\documentstyle[12pt]{article}

\input epsf

\begin{document}
\baselineskip=20.5pt

\def\beqra{\begin{eqnarray}} \def\eeqra{\end{eqnarray}}
\def\beqast{\begin{eqnarray*}}
\def\eeqast{\end{eqnarray*}}
\def\beq{\begin{equation}}      \def\eeq{\end{equation}}
\def\be{\begin{enumerate}}   \def\ee{\end{enumerate}}

%title page
\def\fnote#1#2{\begingroup\def\thefootnote{#1}\footnote{
#2}
\addtocounter
{footnote}{-1}\endgroup}

\def\itp#1#2{\hfill{NSF-ITP-{#1}-{#2}}}

\def\gam{\gamma}
\def\Gam{\Gamma}
\def\la{\lambda}
\def\eps{\epsilon}
\def\La{\Lambda}
\def\si{\sigma}
\def\Si{\Sigma}
\def\al{\alpha}
\def\Tha{\Theta}
\def\tha{\theta}
\def\vphi{\varphi}
\def\del{\delta}
\def\Del{\Delta}
\def\ab{\alpha\beta}
\def\om{\omega}
\def\Om{\Omega}
\def\mn{\mu\nu}
\def\mun{^{\mu}{}_{\nu}}
\def\kap{\kappa}
\def\rsi{\rho\sigma}
\def\beal{\beta\alpha}

\def\til{\tilde}
\def\rta{\rightarrow}
\def\eqv{\equiv}
\def\nab{\nabla}
\def\pa{\partial}
\def\sit{\tilde\sigma}
\def\ul{\underline}
\def\indt{\parindent2.5em}
\def\nd{\noindent}

\def\rsi{\rho\sigma}
\def\beal{\beta\alpha}

        % calligraphic
\def\caa{{\cal A}}
\def\cb{{\cal B}}
\def\cac{{\cal C}}
\def\cd{{\cal D}}
\def\ce{{\cal E}}
\def\cf{{\cal F}}
\def\cg{{\cal G}}
\def\cah{{\cal H}}
\def\ci{{\cal I}}
\def\cj{{\cal{J}}}
\def\ck{{\cal K}}
\def\cl{{\cal L}}
\def\cm{{\cal M}}
\def\cn{{\cal N}}
\def\cO{{\cal O}}
\def\cp{{\cal P}}
\def\car{{\cal R}}
\def\cs{{\cal S}}
\def\ct{{\cal{T}}}
\def\cu{{\cal{U}}}
\def\cv{{\cal{V}}}
\def\cw{{\cal{W}}}
\def\cx{{\cal{X}}}
\def\cy{{\cal{Y}}}
\def\cz{{\cal{Z}}}

        % nots
\def\raisenot{\raise .5mm\hbox{/}}
\def\nota{\ \hbox{{$a$}\kern-.49em\hbox{/}}}
\def\notA{\hbox{{$A$}\kern-.54em\hbox{\raisenot}}}
\def\notb{\ \hbox{{$b$}\kern-.47em\hbox{/}}}
\def\notB{\ \hbox{{$B$}\kern-.60em\hbox{\raisenot}}}
\def\notc{\ \hbox{{$c$}\kern-.45em\hbox{/}}}
\def\notd{\ \hbox{{$d$}\kern-.53em\hbox{/}}}
\def\notbd{\ \hbox{{$D$}\kern-.61em\hbox{\raisenot}}} %big D
\def\note{\ \hbox{{$e$}\kern-.47em\hbox{/}}}
\def\notk{\ \hbox{{$k$}\kern-.51em\hbox{/}}}
\def\notp{\ \hbox{{$p$}\kern-.43em\hbox{/}}}
\def\notq{\ \hbox{{$q$}\kern-.47em\hbox{/}}}
\def\notW{\ \hbox{{$W$}\kern-.75em\hbox{\raisenot}}}
\def\notz{\ \hbox{{$Z$}\kern-.61em\hbox{\raisenot}}}
\def\notpa{\hbox{{$\partial$}\kern-.54em\hbox{\raisenot}}}

\def\fo{\hbox{{1}\kern-.25em\hbox{l}}}  %raised one
\def\rf#1{$^{#1}$}
\def\bx{\Box}
\def\tr{{\rm Tr}}
\def\rmtr{{\rm tr}}
\def\dgg{\dagger}

\def\lag{\langle}
\def\rag{\rangle}
\def\bmid{\big|}

\def\vlap{\overrightarrow{\La p}} %overrightarrow
\def\lrta{\longrightarrow}
\def\lrar{\raisebox{.8ex}{$\longrightarrow$}}
\def\rlarw{\longleftarrow\!\!\!\!\!\!\!\!\!\!\!\lrar}

\def\llra{\relbar\joinrel\longrightarrow}     %THIS IS LONG
\def\mapright#1{\smash{\mathop{\llra}\limits_{#1}}}
\def\mapup#1{\smash{\mathop{\llra}\limits^{#1}}}
\def\asymptotic{{_{\stackrel{\displaystyle\longrightarrow}
{x\rightarrow\pm\infty}}\,\, }} %x goes to plus minus infinity, display sty.
\def\asymptext{\raisebox{.6ex}{${_{\stackrel{\displaystyle\longrightarrow}
{x\rightarrow\pm\infty}}\,\, }$}} %x goes to plus minus infinity,

\def\7#1#2{\mathop{\null#2}\limits^{#1}}   % puts #1 atop #2
\def\5#1#2{\mathop{\null#2}\limits_{#1}}   % puts #1 beneath #2
\def\too#1{\stackrel{#1}{\to}}
\def\tooo#1{\stackrel{#1}{\longleftarrow}}
\def\nout{{\rm in \atop out}}

\def\one{\raisebox{.5ex}{1}}
\def\BM#1{\mbox{\boldmath{$#1$}}}

\def\ltsim{\matrix{<\cr\noalign{\vskip-7pt}\sim\cr}}
\def\gtsim{\matrix{>\cr\noalign{\vskip-7pt}\sim\cr}}
\def\haf{\frac{1}{2}}

%       pictures

\def\place#1#2#3{\vbox to0pt{\kern-\parskip\kern-7pt
                             \kern-#2truein\hbox{\kern#1truein #3}
                             \vss}\nointerlineskip}

\def\illustration #1 by #2 (#3){\vbox to #2{\hrule width #1
height 0pt
depth
0pt
                                       \vfill\special{illustration #3}}}

\def\scaledillustration #1 by #2 (#3 scaled #4){{\dimen0=#1
\dimen1=#2
           \divide\dimen0 by 1000 \multiply\dimen0 by #4
            \divide\dimen1 by 1000 \multiply\dimen1 by #4
            \illustration \dimen0 by \dimen1 (#3 scaled #4)}}

\def\ON{{\cal O}(N)}
\def\UN{{\cal U}(N)}
\def\bdPh{\mbox{\boldmath{$\dot{\!\Phi}$}}}
\def\bPh{\mbox{\boldmath{$\Phi$}}}
\def\bPhs{\bPh^2}
\def\sef{S_{eff}[\sigma,\pi]}
\def\sigx{\sigma(x)}
\def\pix{\pi(x)}
\def\bph{\mbox{\boldmath{$\phi$}}}
\def\bphs{\bph^2}
\def\ex{\BM{x}}
\def\exs{\ex^2}
\def\xdot{\dot{\!\ex}}
\def\y{\BM{y}}
\def\ys{\y^2}
\def\ydot{\dot{\!\y}}
\def\pat{\pa_t}
\def\pax{\pa_x}

\renewcommand{\theequation}{\arabic{equation}}

\itp{96}{127}
%\today

\hfill{hep-th/9610009}\\

\vspace*{.3in}
\begin{center}
 \large{\bf Fermion Bags in the
Massive Gross-Neveu Model}
\normalsize

\vspace{36pt}
Joshua Feinberg\fnote{*}{{\it e-mail: joshua@itp.ucsb.edu}}
 \& A. Zee\\

\vspace{12pt}
 {\small \em Institute for Theoretical Physics,}\\ {\small \em
University of California, Santa Barbara, CA 93106, USA}
\vspace{.6cm}

\end{center}

\begin{minipage}{5.3in}
{\abstract~~~~~

As has long been known, it is energetically favorable for
massive fermions
to deform the homogeneous vacuum around
them, giving rise to extended bag-like objects. We study this
phenomenon
non-perturbatively in a model field theory, the $1+1$
dimensional Massive
Gross-Neveu model, in the large $N$ limit. We prove that
the bags
in this model are necessarily time dependent. We calculate
their masses
variationally and demonstrate
their stability. We find a non-analytic behavior in these masses as we
approach the standard massless Gross-Neveu model and argue that this
behavior is caused by the kink-antikink threshold. This work extends our
previous work to a
non-integrable
field theory.}
\end{minipage}

\vspace{48pt}


\vspace{48pt}

A central concept in particle physics states that
fundamental particles
acquire their masses through interactions with vacuum
condensates. Thus, a
massive particle may carve out around itself a spherical
region
\cite{sphericalbag} or a shell \cite{shellbag} in which the
condensate is
suppressed, thus reducing the effective mass of the particle
at the expense
of volume and gradient energy associated with the
condensate. This picture
has interesting phenomenological consequences
\cite{sphericalbag,
mackenzie}.

Here we study these effects within the $1+1$ dimensional
massive generalization of the Gross-Neveu model \cite{gn}
(which we will
refer to as MGN),
\beqra
S&=&\int d^2x\left\{\sum_{a=1}^N\, \bar\psi_a\,\left(i\notpa
-
M\right)\,\psi_a +
\frac{g^2}{2}\; \left(
\sum_{a=1}^N\;\bar\psi_a\,\psi_a\right)^2\right\}\nonumber\\
&=&\int
d^2x\,\left\{\bar\psi\left[i\notpa-\si\right]\psi-{1\over
2g^2}\left(
\si^2-2M\si\right)\right\}
\label{lagrangian}
\eeqra
describing $N$ self interacting massive Dirac fermions
$\psi_a$ carrying a flavor index $a=1,\ldots,N$, which we
promptly
suppress. As usual, the theory can be rewritten with the
help of  a scalar
flavor singlet auxiliary field $\si(x)$. Also as usual, we take
the large
$N$ limit holding $\lambda\equiv Ng^2$ fixed. Integrating
out the
fermions,
we obtain the bare effective action
\beq
S[\si] =-{1\over 2g^2}\int\, d^2x
\,\left(\si^2-2M\si\right) -iN\,
\tr~{\rm log}\left(i\notpa-\si\right)\,.
\label{fermout}
\eeq
Noting that $\gam_5 ( i\notpa -\si )= -(i\notpa +\si)\gam_5
$,  we can
rewrite the $\tr~{\rm log}(i\notpa -\si )$ as ${1\over 2}
\tr~{\rm log}
(i\notpa -\si
)(i\notpa +\si)$. If $\si$ is time independent, this may be further simplified
to $ {1\over 2}\int {d\om \over 2 \pi} [(\tr~{\rm log} (h_+-
\om^2)+\tr~{\rm log} (h_-
-\om^2)]$ where $h_{\pm}  \equiv -\pa_x^2 + \si^2 \pm \si'
$. Clearly, the
two Schr\"odinger operators $h_{\pm}$ are
isospectral (see Sec. II of \cite{josh1}) and thus we obtain
\beq
S[\si] =-{1\over 2g^2}\int\, d^2x
\,\left(\si^2-2M\si\right) -iN\,
\int\limits_{-\infty}^{\infty} {d\om \over 2 \pi}\tr~{\rm log}(h_- -\om^2)
\label{effective}
\eeq

In contrast to the standard massless Gross-Neveu model
(the GN model),
the
MGN model studied here is not invariant under the $Z_2$
symmetry $\psi
\rightarrow \gamma_5 \psi$, $\sigma \rightarrow -\sigma$,
and the physics
is correspondingly quite different. The GN model contains a
soliton (the so
called CCGZ kink \cite{ccgz, dhn, josh1}) in which the
$\sigma$ field takes
on
equal and opposite values at $x=\pm\infty$. The stability of
this soliton is
obviously guaranteed by topological considerations. With
any non-zero
$M$
the vacuum value of $\sigma$ is unique and the CCGZ kink
becomes
infinitely
massive and disappears. If any soliton exists at all, its
stability has to
depend on the energetics of trapping fermions. Also, the GN
model is
completely integrable, while the MGN model is widely
believed to be
non-integrable. In recent work \cite{feinzee, josh1} we have
studied
integrable models, and one of the purposes of this note is to
show that it
is possible to obtain non-perturbative results even for non-
integrable
models, albeit in the large $N$ limit.

{\em The vacuum state} ~~~Setting $\si$ to a constant we
obtain from
(\ref{effective}) the renormalized effective potential (per
flavor)
\beq\label{veff}
V(\si,\mu) = {\si^2\over 4\pi}~ {\rm log}~ {\si^2\over
e\mu^2} +
{1\over \lambda(\mu)}~\left[{\si^2\over 2} -
M(\mu)\si\right]\,,
\eeq
where $\mu$ is a sliding renormalization scale with
$\lambda(\mu)=Ng^2(\mu)$ and
$M(\mu)$ the running couplings. By equating the coefficient
of $\si^2$ in
two versions of $V$, one defined with $\mu_1$ and the
other with
$\mu_2$,
we find immediately that
\beq\label{scale}
{1\over\lambda(\mu_1)} - {1\over\lambda(\mu_2)} =
{1\over \pi}~{\rm
log}
~{\mu_1\over\mu_2}
\eeq
and thus the coupling $\lambda$ is asymptotically free, just
as in the GN
model. Furthermore, by equating the coefficient of $\sigma$
in $V$ we see
that the ratio ${M(\mu)\over\lambda(\mu)}$
is a renormalization group invariant. Thus, $M$ and
$\lambda$ have the
same
scale dependence.

Without loss of generality we assume that $M(\mu)>0$  and
thus the
absolute
minimum of (\ref{veff}), namely, the condensate
$m=\langle\si\rangle$, is
the positive solution of the gap equation
\beq\label{gap}
{dV\over d\si}~{\Big|_{\si=m}}= m\left[ {1\over \pi}~{\rm log}
~{m\over\mu} + {1\over \lambda(\mu)}\right] -
{M(\mu)\over\lambda(\mu)} = 0\,.
\eeq
Referring to (\ref{lagrangian}), we see that $m$ is the mass
of the
fermion.
Using (\ref{scale}), we can re-write the gap equation as
${m\over\lambda(m)}= {M(\mu)\over\lambda(\mu)}$,
which shows
manifestly
that $m$, an observable physical quantity, is a
renormalization group
invariant.
This equation also implies that $M(m)=m$, which makes
sense physically.

{\em Static space dependent $\sigx$ backgrounds}
~~~Ideally, we would
like
to solve the field equation ${\delta S \over \delta
\sigma(x,t)}=0$, a
difficult task beyond the capability of field theorists at
present. A more
realistic goal is to restrict ourselves to time-independent
$\sigma$ field
and to try to solve ${\delta S \over \delta \sigma(x)}=0$, but
even that is
difficult since we don't know how to evaluate $S$ for an
arbitray time
independent but space dependent $\sigma(x)$.
Furthermore, we can show
(see
below) that such a solution does not exist.

The relevant physics is not difficult to understand. A
generic $\sigx$
will distort the fermion vacuum, causing the fermions to
back-react on
$\sigx$ so as to minimize their energy, and in general $\si$
will become
time dependent. In our previous work \cite{feinzee} we
have found a
necessary condition (albeit generically insufficient) to avoid
such a
back-reaction in $1+1$ dimensional theories.
The condition is physical and easy to state (and we will state
it in the
present context.) Consider the expectation value of the
fermionic vector
current $j^{\mu}(x) $ in the background specified by a field
configuration
$\si$. After some standard manipulations we could show
\cite{feinzee}
that
the spatial component of the fermion number current
will not vanish at spatial infinity, unless the Schr\"odinger
operator $h_-$
is reflectionless.  Moreover, the fermion number current
will run in
opposite directions at $x=\pm\infty$. This apparent current
non-conservation indicates that a state giving rise to a static
reflectionful $h_-$
is highly unstable and will immediately try to decay to a
stable state by
emitting fermions.

We will thus restrict ourselves to only those $\si$
configurations which
correspond to reflectionless $h_-$. Since there are only
denumerably
infinite number of reflectionless Schr\"odinger operators
known, this
condition vastly restrict the space of possible $\sigx$. Our
calculation
amounts to a variational calculation in quantum field
theory. For any given
fermion number $N_f$ the energy of the bag or lump we
calculate below
is an
upper bound to the true energy.

{\em A variational calculation of the bag mass} ~~~This
upper bound on
the true energy cannot be saturated by static $\sigx$
configurations,
because
as we already mentioned, the MGN model does not have
static saddle point
$\sigx$ configurations. However, it is clear from the
discussion above that reflectionless $\sigx$ configurations
are the best trial configuration among
all static configurations. As usual, the art behind a variational calculation
consists of a judicious choice of a trial function.

The energy functional (per flavor) $\ce [\sigx]$ for static
$\sigx$
configurations is by definition $\ce = - {S\over NT}$ where
$T$ is some
temporal infrared cutoff. We write (\ref{effective}) as
\cite{josh1}
\beq\label{stateff}
\ce [\sigx] = {1\over 2\lambda} \int\limits_{-\infty}^{\infty}
dx~[V(x) -2M\sigx] - \int\limits_{-\infty}^{\infty}
{d\om\over 2\pi
i}~\tr~{\rm
log}~[-\pa_x^2 + V(x) -\om^2]
\eeq
where $V(x) = \si^2(x) - \si'(x)$. (Here we used
$\int\limits_{-\infty}^{\infty}dx~\si'(x) =0$ by invoking the
boundary conditions $\sigx\asymptext m$.)

Out of the denumerably infinite number of
reflectionless Schr\"odinger operators we now take the
simplest possibility:
that $h_- =-\pa_x^2 + \si^2 - \si' $ has a single normalizable
bound state at
some positive \cite{positive} energy $\om_b^2<m^2$ (and
thus bound
states at $\pm \om_b$ in the Dirac
operator.) It is well-known from the annals of quantum
mechanics that
these properties uniquely determine the single parameter
family of
potentials
\beq\label{potential}
V(x)= m^2 - 2\kappa^2~{\rm sech}^2~[\kappa(x-x_0)]
\eeq
(up to an overall translation parameter $x_0$ which we
immediately set to
zero.) The normalized bound state wave function is
$\psi_b(x) =
\sqrt{{\kappa\over 2}}~{\rm sech}~\kappa x$. The bound
state energy $\om_b^2$ is
given by $\kappa=\sqrt{m^2-\om_b^2}$, thus suggesting
that we trade
$\kappa$ immediately for an angle ${\pi\over
2}\geq\theta\geq 0$ such that
$\kappa = m~{\rm sin} \theta$ (and thus $\om_b = m~ {\rm
cos}\theta$.)
The corresponding $\sigx$ is:
\beq\label{si}
\sigx = m + \kappa \left[ {\rm tanh }~\left( \kappa x -
{1\over 4}~{\rm
log}~{m+\kappa\over m-\kappa}\right) - {\rm tanh }~\left(
\kappa x +
{1\over 4}~{\rm log}~{m+\kappa\over m-\kappa}\right)
\right]\,.
\eeq

With (\ref{si}) as a trial configuration, the energy
(\ref{stateff}) becomes
an ordinary function $\ce (\theta)$. We thus vary with
respect to the
variational parameter $\theta$ (or equivalently $\kappa$.)
The extremum condition on the energy is
\beq\label{extremum}
{\pa \ce\over \pa \theta} = \int\limits_{-\infty}^{\infty}
~dx~\left\{ \left[
{1\over 2\lambda} -
\int\limits_{-\infty}^{\infty}~{d\om\over 2\pi
i}~R(x,\om)\right]~{\pa
V\over \pa \theta} - {M\over \lambda}~{\pa\si\over
\pa\theta}~\right\}\,=0\,.
\eeq
Here $R(x,\om) \equiv \langle x |(-\pa_x^2 + \si^2 - \si'
-\om^2 )^{-1}| x\rangle $ denotes the resolvent of $h_-$,
and can be
calculated to be
\cite{josh1}
\beq\label{resolvent}
 R(x,\om) = {1\over 2 \sqrt{m^2-\om^2}}~\left[ 1+ {m^2-
\si^2+\si'\over
2(\om_b^2-\om^2)}\right] =
{1\over 2 \sqrt{m^2-\om^2}}~\left[ 1+ {2\kappa
\psi_b^2(x)\over
\om_b^2-\om^2} \right]\,.
\eeq
Substituting (\ref{resolvent}) in (\ref{extremum}) we find
\beqra\label{extremum1}
{\pa \ce\over \pa \theta} &=& \int\limits_{-\infty}^{\infty}
~dx~\left\{
{1\over 2}\left[{1\over \lambda} - \int\limits_{-
\infty}^{\infty}~{d\om\over 2\pi i}~
{1\over \sqrt{m^2-\om^2}}\right]~{\pa V\over \pa \theta} -
{M\over
\lambda}~{\pa\si\over \pa\theta}~\right\}\nonumber\\ &-
&
\kappa~I(\om_b, m)
~\langle\psi_b\Big|
~{\pa V\over \pa\theta}~\Big|\psi_b\rangle\,,
\eeqra
where $I(\om_b,m)\equiv\int_{\cal C} {d\om\over 2\pi i}
{1\over
(\om_b^2-\om^2)~\sqrt{m^2-\om^2}}$ (the contour ${\cal
C}$ is specified
below.)

Note that the $\om$ integral in the first term in
(\ref{extremum1}) is
logarithmically divergent. This UV divergence is taken care
of as follows.
Let us consider (\ref{extremum1}) at the cutoff scale
$\Lambda$. Setting
$\si$ to its vacuum value in $\delta\ce /\delta\sigx = 0$ (see
(\ref{staticsaddle}) below)
we have the
(bare) gap equation
\beq\label{baregap}
{1- { M(\Lambda)\over m}\over \lambda(\Lambda)} =
\int\limits^{\Lambda}_{-\Lambda}\, {d\om \over 2\pi}
{1\over \sqrt{\om^2 -m^2 +i\epsilon}}\,.
\eeq
Using (\ref{baregap}) in (\ref{extremum1}) we see that all
reference to
$\Lambda$ disappears and the extremum condition becomes
\beq\label{nigzeret}
{\pa\ce\over\pa\theta} = {M\over 2\lambda m}~{\pa\over
\pa \theta}
\int\limits_{-\infty}^{\infty} dx~[(\si-m)^2-\si'] - \kappa~I
(\om_b,m)
~\langle\psi_b\Big|
~{\pa V\over \pa\theta}~\Big|\psi_b\rangle\,.
\eeq
To evaluate the integral $I(\om_b,m)$, we have to choose the
proper contour
${\cal C}$, and thus
we have to invoke our understanding of the physics of fermions. We fill
the
Dirac sea, including the discrete state at $-\om_b$, and then
put $N_f$
fermions into the state at $\om_b$. Mathematically, we thus
have to let
${\cal C}$ enclose the cut on the
negative $\om$ axis and then go around the pole at $-
\om_b$ $N$ times
and
around the pole at $\om_b$ $N_f$ times. In
this way, we obtain \cite{josh1, dhn} $I={({2\theta\over \pi}
-
\nu)/m^2~{\rm
sin} 2\theta}$ where we have introduced the ``filling
fraction"
$\nu={N_f\over N}$.

Recalling first order perturbation theory we immediately
recognize the
matrix-element in (\ref{nigzeret}) as simply
$\pa \om_b^2 /\pa\theta$.
Putting it all together we find the extremum condition
\beq\label{minimum}
{\pa\ce\over\pa\theta} = 2m~\left[ \left({\theta\over \pi} -
{\nu\over
2}\right)  + \gamma~{\rm tan}\theta\right]~{\rm
sin}\theta\,=0\,.
\eeq
(where we have defined the renormalization group
invariant ratio
$\gamma \equiv {M\over \lambda m}$), thus fixing $\theta$
as a function of
the filling fraction
\beq\label{theta}
{\theta\over \pi} +\gamma~{\rm tan}\theta = {\nu\over
2}\,.
\eeq
Integrating (\ref{nigzeret}) and using (\ref{theta}) we
find that the mass $M$ (namely, $N\ce$ evaluated at the
extremal point) of
our
bag or lump is
\beq\label{mass}
{M (\nu, \gamma)\over Nm} = {2\over \pi}~{\rm sin} \theta
+
\gamma~{\rm
log}~ {1+{\rm sin} \theta\over 1-{\rm sin} \theta}\,.
\eeq

By calculating ${d^2 \ce\over d\nu^2} = -\pi {\rm sin} \theta
/(1 +
\pi\gamma~{\rm sec}^2 \theta)$ we see that $\ce(\nu)$ is a
convex function
and thus satisfies $\ce(\nu_1 + \nu_2) < \ce(\nu_1) +
\ce(\nu_2)$.
Therefore, a lump binding $N\nu$ fermions is variationally
stable against
decaying into two lumps with $N\nu_1$ and  $N\nu_2$
fermions
respectively (with
$\nu=\nu_1+\nu_2 <1$.) Thus, the lump binding $N\nu$
fermions is the
most variationally stable static configuration at the sector of
fermion
number $N\nu$. Note that this is true for small as well as for
large values
of $\gamma$. Furthermore, it is clear from (\ref{mass}) that
the binding
energy (in units
of $m$)
per fermion $B(\nu,\gamma) = 1 - {M(\nu,\gamma)\over
m\nu}$ increases with $\nu$, and does not saturate as
in nuclear
physics.
This is characteristic of the bag picture, in which each
additional particle
digs a deeper hole in the vacuum condensate (``the
mattress effect"). To demonstrate these facts we present
in Fig. (1)
a numerical computation of the binding energy per fermion
at a particular $\gamma$.
%\begin{figure}
%\epsfysize=4 truein
%\epsfxsize=4 truein
%\centerline{\epsffile{bepf.ps}}
%\vspace{-0.1 truein}
%\caption[]{ The binding energy per fermion $B(\nu,
%\gamma)$ at $\gamma=0.1$}
%\end{figure}
%%%%%%%
\vspace{25pt}
\begin{center}   
\epsfxsize 4.0in \epsfbox{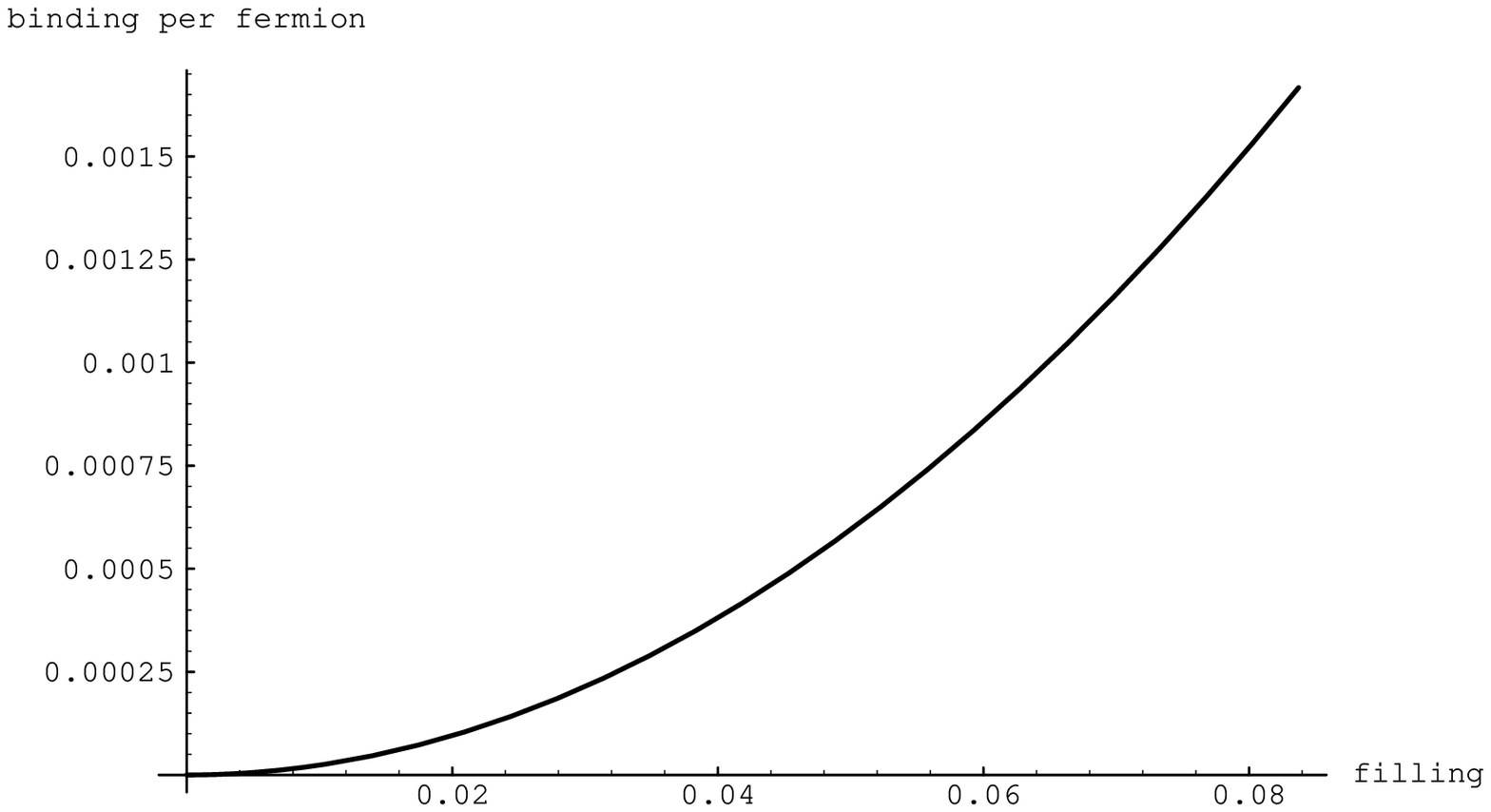}
\end{center}
\vspace{24pt}
{\footnotesize 
Figure 1: The binding energy per fermion $B(\nu, \gamma)$ at $\gamma=0.1$}
\vspace{25pt}
%%%%%%%%%%%%%%%%%%%%

Let us consider the
most stable bag, namely the bag at $\nu=1$. In
the small $\gamma$  limit,
\beq\label{masssmallgamma}
{M (1,\gamma)\over Nm}\sim {2\over \pi}  - \gamma {\rm
log} {\pi e\gamma\over 4} +{\cal
O}(\gamma^{3\over 2})\,.
\eeq
Thus $M(1,\gamma)$ is non-analytic at $\gamma=0$, {\em
i.e.,} at the GN
point. Note that $M(1,0) = {2Nm\over \pi}$ is the kink-antikink threshold
of the GN model \cite{dhn}. It
appears that the
logarithmic
singularity in $M(1,\gamma)$ is associated with the
enhanced $Z_2$
symmetry at
$\gamma=0$. To further argue in this direction, we find that
as soon
as $\nu$ decreases from $1$, which puts us below the kink-
antikink
threshold, $M(\nu,\gamma) =
{2\over \pi} {\rm cos} {\pi (1-\nu)\over 2} + [{\rm log}~({1+ {\rm cos}{\pi (1-\nu)\over
2} \over 1- {\rm cos} {\pi (1-\nu)\over 2}}) -({4\over\pi (1-\nu)}-{\pi (1-\nu)
\over 3}) {\rm sin} {\pi (1-\nu)\over 2}] \gamma + {\cal O}(\gamma^2)
= {2\over \pi} - 2 \gamma~ {\rm log}~ {\pi e (1-\nu)\over 4}
+{\cal O}((1-\nu)^2, \gamma^2, (1-\nu)^2\gamma)\,.$
Thus, the logarithmic singularity $\gamma {\rm log}\gamma$ as
$\gamma\rightarrow 0$
is replaced by a logarithmic singularity $\gamma {\rm log}(1-\nu)$ as
$\nu\rightarrow 1$.
This means that the kink-antikink is indeed the source of
this singular behavior. It would be interesting to address
this
issue in the framework of an appropriate effective action for
bags.

We now turn to the large $\gamma$ limit, which may be
attained by
making the four-fermi interactions weak. The theory should
then describe
quasi-free heavy fermions of mass $m$. We thus expect that
the binding
energy of bags will tend to zero as
$\gamma\rightarrow\infty$. This is
indeed the
case, and we find ${M(\nu,\gamma)\over m}\sim \nu
-{1\over 24}
({\nu^3\over\gamma^2})+{\cal O}(\gamma^{-3})$. The
$\nu^3$ behavior
is once again a manifestation of the mattress effect.
We present the results of the numerical computation of the
binding energy
per fermion at maximal filling $B(1,\gamma)$ in Fig. (2).
%\begin{figure}
%\epsfysize=4 truein
%\epsfxsize=4 truein
%\centerline{\epsffile{be.ps}}
%\vspace{-0.1 truein}
%\caption[]{ The binding energy per fermion $B(1, \gamma)$
%at maximal filling.}
%\end{figure}
%%%%%%%
\vspace{25pt}
\begin{center}   
\epsfxsize 4.0in \epsfbox{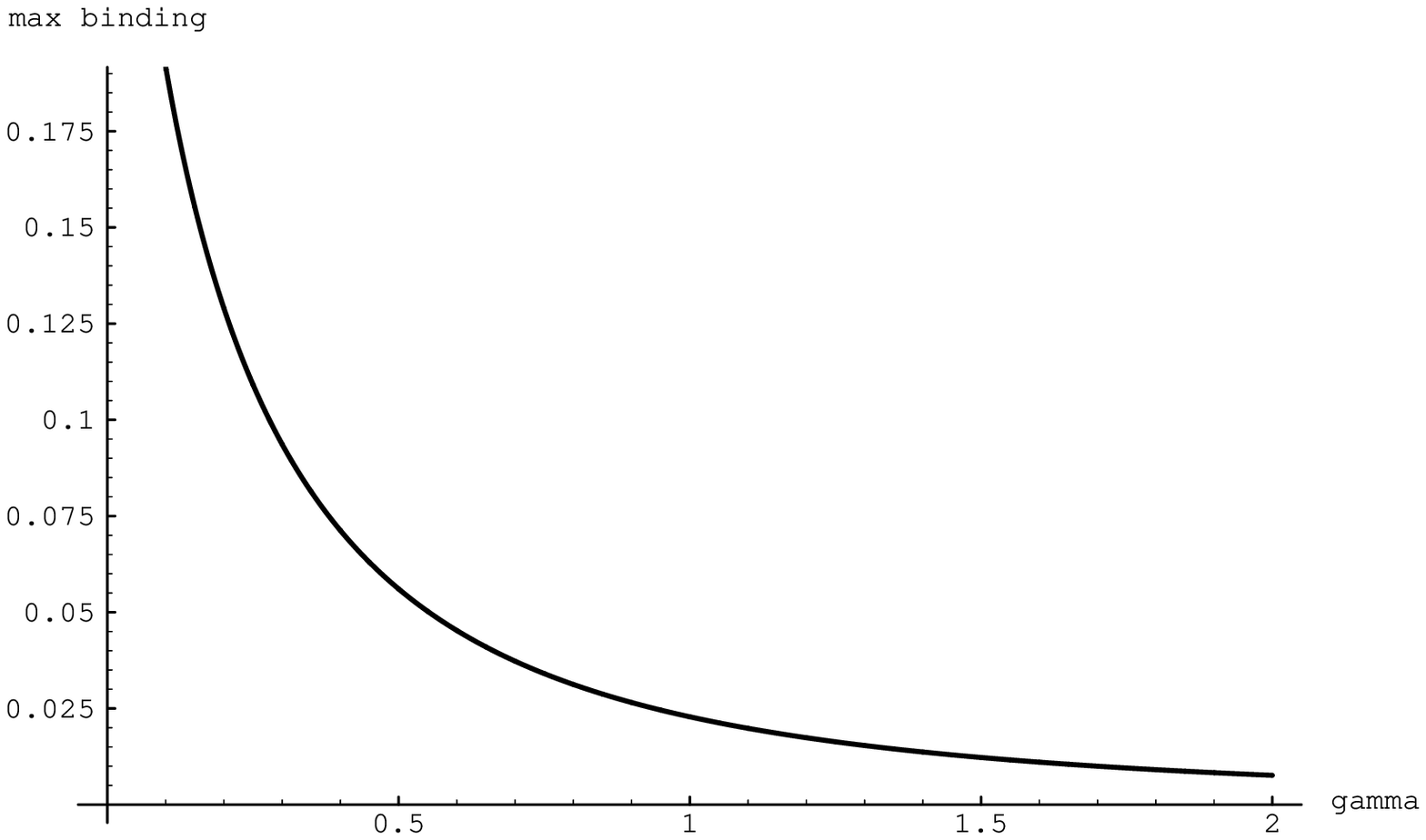}
\end{center}
\vspace{24pt}
{\footnotesize 
Figure 2: The binding energy per fermion $B(1, \gamma)$ at maximal filling.}
\vspace{25pt}
%%%%%%%%%%%%%%%%%%%%

{\em Static bags do not exist}
~~~The extremum condition $\delta\ce /\delta\sigx = 0$ reads
\beq\label{staticsaddle}
i~{\sigx - M\over \lambda} =[2\sigx +
\pa_x]~\int\limits_{-\infty}^{\infty}~{d\om\over
2\pi}~\langle x\Big |~
{1\over
-\pa_x^2 + \si^2 - \si' - \om^2 }~\Big | x \rangle\,.
\eeq
A static saddle point configuration $\sigx$ is necessarily
reflectionless. Concentrating on the case of a single bound state and
substituting
(\ref{resolvent}) into (\ref{staticsaddle}) and using
(\ref{baregap}), we
find
\beq\label{inconsistent}
{M(\mu)\over m\lambda(\mu)}~(\si-m) = (2\si +
\pa_x)~(m^2 - \si^2 +
\si')(-i I(\om_b,m)/4)\,.
\eeq
Further substituting $\sigx$ as given in (\ref{si}) into this
equation, we
verify easily that there is no combination $\kappa
(m,M,\lambda)$ for
which (\ref{inconsistent}) is satisfied. Thus
(\ref{lagrangian}) does not
have
a static reflectionless $\sigx$ saddle point with a single
bound state. (The
possibility that it has a static reflectionless saddle point with
more than a
single bound state seems highly unlikely, but we have not
ruled it out
rigorously.) We thus conclude that
the MGN model does not have any static $\sigx$
configurations. This is in
contrast to the GN model, which has, as we have already
mentioned, static
topological $\sigx$ configurations, all of them are
reflectionless, with a
single bound state \cite{josh1, ccgz, dhn}, or more
\cite{feinzee1}.

The time dependent bags in the MGN model may be seen as
continuous
vibrating deformations of the static non-topological bags of
the GN model
(at least for small values of $\gamma$.) Indeed, the bag
configuration in
the GN model with one bound state is also described by the
$\si$ field
given in (\ref{si}). Dashen et al \cite{dhn} showed long ago
that $\kappa$
and $m$ are quantized according to (\ref{theta}) and
(\ref{mass}), with
$\gamma$ set to zero \cite{josh1, ccgz}. These quantization
conditions, as
well as the gap equation (\ref{gap}), are continuous at
$\gamma=0$. Thus,
this configuration should not change abruptly as we turn $M$
on, but it cannot
remain static. This means that as we switch on $\gamma$,
$\sigx$ will start
vibrating around the static profile (\ref{si}).  The vibration
amplitude and
frequency of these objects must be continuous functions of
$\gamma$ that
vanish as $\gamma\rightarrow 0$. It would be interesting
to
determine whether and how the lack of static $\sigx$ saddle
points in the
massive GN model is related to the common lore that
turning $M$ on
destroys the
 complete integrability of the GN model.

{\em  The fermion current operator and the bosonized
theory}
~~~By using the methods of Sec. 2 of \cite{feinzee} we easily
obtain the
expectation value of the conserved fermion current
$j^{\mu}=\bar\psi\gam^{\mu}\psi$
in the background of a static extremal $\sigx$ configuration
trapping
$N\nu$
fermions. The spatial component is identically zero and the
fermion density
$\langle \si\Big| j^0 \Big|\si\rangle $ is
\beq\label{expectation}
\rho (x) = {N\nu\over 4\kappa}~(\si^2-m^2)\,,
\eeq
which has the correct normalization $\int\limits_{-
\infty}^{\infty} dx~\rho
(x) = N\nu$, as can be seen from (\ref{si}). This means that,
in the
bosonized
theory, the flavor singlet boson $\phi$ develops a spatially
varying
profile which follows the profiles of $\sigx$ according to
\beq
\pa_x \phi (x) =\sqrt{\pi\over 8}~ {\nu\over\kappa}~(\si^2-
m^2)\,.
\eeq

{\bf Acknowledgements}~~~ We are grateful to Mark
Srednicki for
indispensable help in carrying out the numerical work
behind the figures.
This work was partly supported by the National Science
Foundation under
Grant No. PHY89-04035.


\begin{thebibliography}{99}
\bibitem{feinzee}  J. Feinberg and A. Zee, preprint cond-
mat/9603173.
~Reflectionlessness of the Dirac operator as a necessary
condition is
discussed
in Section 2. Here we sketch how this comes about. Fermion
bilinears such as the current can be determined from the
resolvent $R(x,\om)$ defined in the text, which can in turn
be determined in terms of the Jost functions  of the
Hamiltonian $h_-$.
\bibitem{positive}  By standard arguments, if
$h_- \psi=0$  had a normalizable solution $\psi$, that solution
would be
$\psi={\rm exp}~[-\int\limits_{x_0}^x \si(y)~dy].$ However, if
$\sigx\asymptext m$, $\psi$ is not square integrable.
\end{thebibliography}
\end{document}